\newcommand{\be}{\begin{equation}}
\newcommand{\ee}{\end{equation}}
\newcommand{\ber}{\begin{eqnarray}}
\newcommand{\eer}{\end{eqnarray}}
\newcommand{\gsim}{\, \raisebox{-0.8ex}{$\stackrel{\textstyle >}{\sim}$ }}
\newcommand{\lsim}{\, \, \raisebox{-0.8ex}{$\stackrel{\textstyle <}{\sim}$ }}
\newcommand{\mbf}{\mathbf}

\def\fm3{fm$^{-3}$}
\def\ms{M_s}

\documentstyle[prl,twocolumn,aps,epsfig]{revtex}

\begin{document}
\title{
Neutrino and axion emissivities of neutron stars from nucleon-nucleon 
scattering data}
\author {Christoph Hanhart$^{1,2}$, 
Daniel R. Phillips$^1$,  
and Sanjay Reddy$^2$
}
\address{$^1$Department of Physics, University of Washington, Box 351560, 
Seattle, WA 98195-1560. \\
$^2$Institute for Nuclear Theory, University of Washington, Box 351550,
Seattle, WA 98195-1550. }
\date{\today}
\maketitle

\begin{abstract} 
Neutrino and axion production in neutron stars occurs mainly as
bremsstrahlung from nucleon-nucleon ($NN$) scattering. The energy
radiated via neutrinos or axions is typically very small compared to
other scales in the two-nucleon system. The rate of emission of such
``soft" radiation is directly related to the on-shell $NN$ amplitude,
and thereby to the $NN$ experimental data. This facilitates the
model-independent calculation of the neutrino and axion radiation
rates which is presented here. We find that the resultant rates are
roughly a factor of four below earlier estimates based on a
one-pion-exchange $NN$ amplitude.
\end{abstract}

Neutron stars are believed to be born during a supernova
explosion with an interior temperature $T$ of order 60 MeV. The
subsequent evolution of the hot and dense compact star is characterized by a
rapid early cooling phase followed by a, significantly slower, late-time
cooling phase~\cite{CG}. During both of these phases 
neutrinos are an important source of energy loss.  Thermal 
evolution of neutron stars is
largely driven by neutrino bremsstrahlung reactions such as

\begin{equation}
NN \rightarrow NN \nu\bar{\nu} \quad \quad nn \rightarrow np e^- \bar{\nu_e}.
\label{eq:nubrem}
\end{equation}

In the first part of the nascent neutron star's life it evolves by
diffusion of trapped neutrinos. This cools the interior to $T \sim 1$
MeV on a time scale of a few seconds~\cite{BL}. The resultant intense
neutrino emission is thought to play a central role in both the
supernova mechanism and r-process nucleosynthesis.  The high neutrino
luminosity is fueled by the reactions (\ref{eq:nubrem}), which compete
with the annihilation $e^++ e^- \rightarrow \nu\bar{\nu}$ in
degenerate matter. At later times, the neutron star enters a period of
slower thermal evolution, during which the emitted neutrinos free
stream. This occurs because the neutrino mean-free path becomes long
when $T \lsim 1$ MeV. The time scale for this long-term cooling of the
dense, degenerate, neutron-rich, inner core thus depends crucially on
the neutrino emissivity, which is, again, dominated by the reactions
(\ref{eq:nubrem}). Observational constraints on this late-time portion
of the neutron star's evolution will improve as X-ray observatories
such as Einstein, EXOSAT and ROSAT \cite{CG} gather pulsar data which
gives information on the surface temperatures of these stars. The
challenge for theorists is to improve the models of both the early-
and late-time cooling of the neutron star.

The emissivities which are key ingredients in these simulations are
dominated by the reactions (\ref{eq:nubrem}).  Despite their central
role in neutron-star dynamics these reactions have received relatively
little attention.  Pioneering work was done by Friman and
Maxwell~\cite{FM} who computed the reaction rates using a
nucleon-nucleon amplitude due only to a single pion exchange
(henceforth we refer to such calculations as the ``OPE
approximation").  Recently, many-body effects, in particular the
suppression due to multiple scattering (the Landau-Pomeranchuk-Migdal,
or LPM, effect \cite{LPM}), have been shown to be important at
temperatures $T \gsim 5$ MeV \cite{KV,RS,SD,Keil}.  However, the work
of Ref.~\cite{FM} is still the state of the art 
treatment of the $NN$ interactions which occur during the
reactions~(\ref{eq:nubrem})~(see also Ref.~\cite{EM}). In this
letter we follow an approach reminiscent of that used in soft-photon
calculations~\cite{L}, and relate the rate of production of
soft-neutrino radiation in $NN$ scattering to the on-shell $NN$
scattering amplitude. This yields a calculation we present as a
``benchmark", which accounts for the two-nucleon dynamics in a
model-independent way. We make no attempt to account for many-body
effects, although they are undoubtedly important in the star. We
identify the density and temperature range over which our results are
valid and show that, although limited, it is of interest to both
supernova and neutron star physics.  In contrast, the full evaluation
of the rates for the reactions (\ref{eq:nubrem}) in a strongly-coupled
medium is a complicated problem. Therefore, of necessity, most
solutions to it will be model-dependent. Thus, we see our results as
providing a model-independent foundation on which future work that
assesses the role of many-body effects can build.

Our main focus in this work will be the neutrino emissivity from $NN
\rightarrow NN \nu \bar{\nu}$.  However, axions (if they exist)
couple, like neutrinos, to the nucleon spin. Therefore, in computing
the neutrino emissivity one obtains the axion emissivity from $NN
\rightarrow NN a$ as a welcome by-product~\cite{RS}.  This is useful
because one important constraint on the axion mass comes from
considering the role of axionstrahlung 
in the dynamics of SN1987A~\cite{Keil,BT}. Indeed, if the rate for the
reaction $NN \rightarrow NN a$ is too high then 
the supernova dynamics is completely changed, and the successful ``standard"
picture of the supernova is destroyed.
Thus, one can constrain the axion coupling,
and hence the axion mass, by demanding that axion radiation did not
make too large a contribution to the energy loss from SN1987A.

\begin{figure}[h,t,b]
\centering
{\epsfig{figure=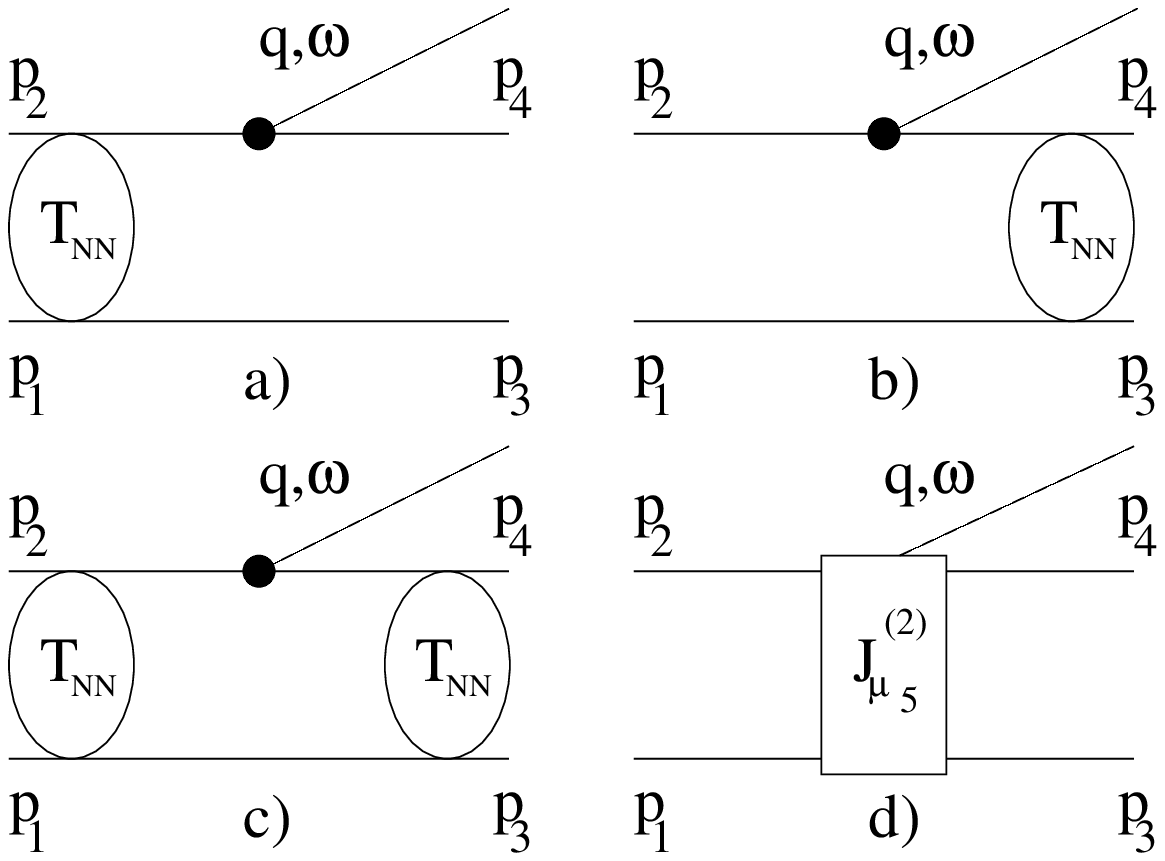,height=.2\textheight}}
\vskip 2 mm
\caption[]{Feynman diagrams for the bremsstrahlung process. The radiation is
represented by the dashed line, and nucleons by solid lines.
{\sc T}$_{NN}$ is the $NN$ transition matrix and $J^{(2)}_{\mu 5}$ 
is a two-body axial current.}
\label{feyn}
\end{figure}

{\it Neutrino and axion emissivities:} We begin by explicitly
calculating the emissivity due to $NN \rightarrow NN \nu
\bar{\nu}$. The $\nu \bar{\nu}$ coupling to non-relativistic baryons
at low energies is given by the Lagrange density 
\be 
{\cal L}_W=-\frac{G_F}{2\sqrt{2}}~ l^\mu~N^\dagger\left (c_v \delta_{\mu,0} -
c_a \delta_{\mu,i}\sigma_i\right)N \,, 
\label{lweak} 
\ee 
where $\l^\mu=\bar{\nu}\gamma^\mu(1-\gamma^5)\nu$ is the leptonic current,
$G_F=1.166 \times 10^{-5}$ GeV$^{-2}$, $N$ is the nucleon field, and
$c_v$ and $c_a$ are the nucleon neutral-current vector and
axial-vector coupling constants.  Some Feynman diagrams for the
bremsstrahlung process are shown in Fig.~\ref{feyn}.

\vskip -2 mm

The incoming (outgoing) nucleon momenta are labeled $\mbf{p_1,p_2}$
($\mbf{p_3,p_4}$). The dashed line represents radiation---a
neutrino-anti-neutrino pair in this case---which carries energy
$\omega$ and momentum $\mbf{q}$. In general we are interested in cases
where the radiated energy is small compared to the incoming nucleon
energy. In the limit $\omega \rightarrow 0$ the amplitudes
corresponding to diagrams (a) and (b) in Fig. 1 are dominant, as they
contain pieces proportional to $1/\omega$. On the other hand, the
contributions from the re-scattering diagram Fig.~1(c), and from
meson-exchange currents such as Fig.~1(d), remain finite in the
$\omega \rightarrow 0$ limit. Thus, for the reaction $nn\rightarrow nn
\nu\bar{\nu}$ the matrix element can be written as

\begin{equation} 
{\cal M}=2 \, \, \frac{G_F}{2\sqrt{2}}~\frac{1}{\omega}~l^\mu \langle 
{\bf p}'|[{\bf T}_{NN} \,, \Gamma_\mu]|{\bf p} \rangle + 
O(\omega^0) \,,
\label{amp1} 
\end{equation} 
where ${\bf p}\,(\mbf{p'})$ is the initial
(final) relative momentum of the two-nucleon system.  We
refer to results which retain only this leading term, of
$O(\omega^{-1})$, in ${\cal M}$ as ``true in the soft-neutrino
approximation (SNA)". In general the $NN$ T-matrix appearing in
Eq.~(\ref{amp1}), ${\bf T}_{NN}$, will be half off-shell~\footnote{As
used here, it should involve a sum over the allowed
partial-waves of the $NN$ system. This, together with the factor of
two in front of the matrix element, accounts for the exchange graphs
which must be included in ${\cal M}$.}. But, in the SNA we can
take ${\bf T}_{NN}$ to be the on-shell $NN$ amplitude.  We can also
neglect the difference between the magnitude of the initial and
final-state relative momenta. We expect these approximations to break
down when $\omega \sim m_\pi$, since $m_\pi$ sets the scale for variations
of {\bf T}$_{NN}$ in the off-shell direction~\footnote{At very low
relative momenta the scale of breakdown is set by the $NN$ scattering length,
since that gives the variation in the on-shell direction. However,
$a_{NN}$ does not really play a role here, since typical nucleon
momenta in neutron stars are at least 100 MeV.}.  So, in the SNA, the
$NN$ interaction is described by the on-shell T-matrix ${\bf
T}_{NN}$, evaluated at a center-of-mass energy which, for reasons of
symmetry, is chosen to be $(p^2 + p'^2)/(2 M)$ ($M$ is the nucleon
mass).  This T-matrix can be constructed from phase shifts deduced
from $NN$ scattering data~\cite{SAID}.  Note that the OPE
approximation used in most previous calculations involves
substituting $V_{OPE}$, the one-pion-exchange {\it potential}, for
${\bf T}_{NN}$ in Eq.~(\ref{amp1}). Meanwhile, $\Gamma_\mu$ is the
vertex which couples the radiation to the nucleons. For $\nu
\bar{\nu}$ radiation $\Gamma_\mu$ follows straight from
Eq.~(\ref{lweak}). Only its three-vector part contributes to ${\cal
M}$ at $O(\omega^{-1})$.  Equation (\ref{amp1}) then gives us a
model-independent result for ${\cal M}$, which is correct in the SNA.

If only two-body collisions are taken into account then the neutrino 
emissivity from a neutron gas is given by Fermi's golden rule
\ber
&& {\cal E}_{\nu\bar{\nu}}=
\int\frac{d^3q_1}{(2\pi)^3 2\omega_1} \frac{d^3q_2}{(2\pi)^3 2\omega_2} 
(2\pi)^4 \delta(E_{in}-E_{fn}) \nonumber \\
&& \quad \omega~\delta^3({\bf p}_{in}-{\bf p}_{fn})
\int \left[\prod_{i=1..4} \frac{d^3p_i}{(2\pi)^3}\right]  {\cal F}~
\frac{1}{s}~\sum_{\rm spin} |{\cal M}|^2 \,,
\label{emiss}
\eer
where ${\cal F}=f_1f_2(1-f_3)(1-f_4)$, with
$f_i=1/(1+\exp{(E_i-\mu_i)/T)}$ being the Fermi-Dirac distribution
function for the nucleons, and $s=4$ the symmetry factor accounting
for identical nucleons. The spin-summed square of the matrix element
can be factored into leptonic and hadronic tensors, and then
represented by
\be
\sum_{\rm spin} |{\cal M}|^2 =\frac{G_F^2c_a^2}{8}~{\sc Tr}~(l^i 
l^j)~{\cal H}_{i,j} \,.
\label{eq:Hdefn}
\ee
The trace over the lepton tensor is easily evaluated. Further, since we are
interested in soft radiation, we may safely ignore $\vec{q}$ in the momentum
delta function~\cite{FM}.  This allows us to directly integrate the leptonic
trace over neutrino angles to obtain
\be
\int d\Omega_1\int d\Omega_2 {\sc Tr}~(l_il_j) = 8~(4\pi)^2 \omega_1\omega_2 ~
\delta_{i,j} \,.
\label{lepint}
\ee
Therefore, only the trace of the 
hadronic tensor ${\cal H}_{ij}$ contributes to the
emissivity, and so we define a scalar function,
\ber
S_{\sigma}(\omega)&=& \int \left[\prod_{i=1..4}\frac{d^3p_i}{(2\pi)^3} 
\right]~(2\pi)^4 \delta^3({\bf p_1+p_2-p_3-p_4})
\nonumber \\  
&&\qquad \delta(E_1+E_2-E_3-E_4-\omega)~{\cal F}~\frac{1}{s}~{\cal H}_{ii} \, ,
\label{ss}
\eer
which is called the dynamical spin structure function of the medium.
It is related to the $\nu \bar{\nu}$ emissivity via:
\begin{equation}
{\cal E}_{\nu\bar{\nu}}=
\frac{G_F^2 c_a^2}{16 \pi^4}\frac{1}{30}\int~d\omega~\omega^6~S_{\sigma}
(\omega) \, ,
\label{enu}
\ee
where $\omega$ is the total energy of the emitted $\nu \bar{\nu}$ pair.

In the two-body approximation considered here we evaluate ${\cal
H}_{ii}$ using Eqs.~(\ref{amp1}) and (\ref{eq:Hdefn}).  For the case
of emission from the $nn$ system, only the spin-triplet two-nucleon state
contributes, and the trace is: 
\be
{\cal H}_{ii}=16 \, \,
\frac{1}{\omega^2}~\sum_{\ms\ms'} \left| \langle 1 \ms',{\bf
p}'|\left[S_i,{\bf T}_{NN}\right]|{\bf p},1 \ms \rangle \right|^2,
\label{hii}
\ee 
where $S_i$ is the total spin of the two-nucleon system. 
It is straightforward to generalize this formula for ${\cal H}_{ii}$ 
to the $np$ case, although the $NN$ spin singlet then 
contributes. (The $np$ case, and the failure of the OPE approximation 
there, is discussed in Ref.~\cite{Sigl}.) From Eqs.~(\ref{hii}) and (\ref{ss})
we can calculate $S_\sigma$, and thus the $\nu \bar{\nu}$ emissivity.

The emission of any radiation which couples to the nucleon spin will
be described by the same function $S_\sigma$. Thus, as mentioned
above, with $S_\sigma$ in hand we may derive the axion emissivity
${\cal E}_a$. The effective theory for axion-nucleon
interactions is described by the Lagrange density ${\cal L}_{\rm ann}
= - g_{\rm ann}~a\bar{N}\gamma^5N$, where 
$a$ is the axion field, and $g_{\rm ann}=10^{-8} (m_a/1 \, {\rm eV})$ 
is the effective axion coupling ($m_a$ is the axion mass)~\cite{DK}.
The calculation of the axion emissivity in this
effective theory is analogous to the above calculation of the
neutrino emissivity, and yields
\be 
{\cal E}_{a}=\frac{g_{\rm
ann}^2}{16 \pi^2 M^2} \frac{1}{3} \int~d\omega~\omega^4~S_\sigma(\omega)\,.
\label{ea} 
\ee 

Before proceeding to our results we note that $S_\sigma$ can be
defined in a much more general way, where it describes the response of
a many-body system to an external spin-dependent
perturbation. Equations~(\ref{enu}) and (\ref{ea}) remain true if this
definition is adopted. To obtain Eq.~(\ref{ss}) for $S_\sigma$ in this
general case one takes the long-wavelength limit of the leading term
in the density expansion.

{\it Results \& Discussion:} We present results for the dynamic spin
structure function $S_{\sigma}(\omega)$, since it includes the
density, temperature, and nuclear dynamics dependence of the neutrino and
axion emissivities. During the evolution of neutron stars, one
encounters varying degrees of nucleon degeneracy, with $\mu_n/T \sim
1$ at birth, but $\mu_n/T \gg 10$ at late times. Earlier
investigations have shown that analytic approximations to the
phase-space integrals in Eq.~(\ref{ss}) work poorly at intermediate
degeneracy~\cite{BT}. Therefore, in this work these integrals are all
performed numerically.  In order to investigate the effect of our
model-independent treatment of the $NN$ interaction we plot the ratio
$R_{\sigma}(\omega) \equiv S^{SNA}_{\sigma}(\omega)/S^{{\rm
ref}}_{\sigma}(\omega)$, where $S^{SNA}_{\sigma}(\omega)$ is
calculated as described above. The denominator, $S^{{\rm
ref}}_{\sigma}(\omega)$, is the dynamic spin structure function found
when a hadronic tensor trace of the form ${\cal H}_{ii}=c/\omega^2$ is
inserted into Eq.~(\ref{ss}). We adjust the constant $c$ so that when
$S_\sigma^{\rm ref}$ is employed in Eq.~(\ref{enu}) the neutrino
emissivity thereby obtained is equal to that found if the {\it full}
OPE matrix element is used in evaluating ${\cal H}_{ii}$~\footnote{We
could have followed Refs.~\cite{Keil,BT} and adopted a reference
$S_\sigma(\omega)$ in which the full OPE-approximation matrix element is
replaced by its value in the $m_\pi \rightarrow 0$ limit. However,
this is a poor approximation to the actual result for one-pion
exchange, since it over-estimates the OPE-approximation emissivities by as
much as a factor of two (see also Refs.~\cite{RS,BT}).}.

\begin{figure}[h,t,b] \centering
{\epsfig{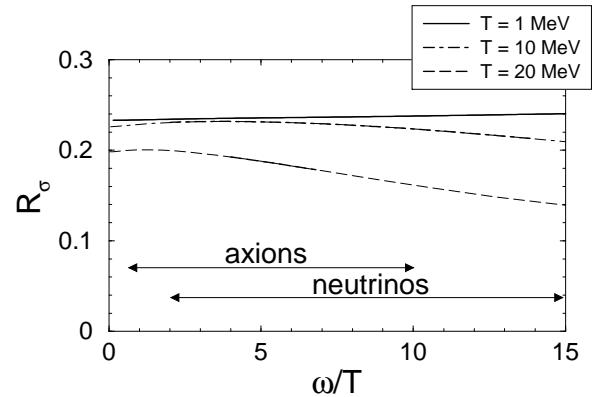}}
\vskip 2 mm
\caption[]{A plot showing the ratio
$R_{\sigma}=S_\sigma^{SNA}/S_\sigma^{\rm ref}$ for $nn$ pairs
in neutron matter at a density of 0.16 fm$^{-3}$. The region where 
multiple scattering  suppression due to the LPM effect is negligible 
($\omega > \gamma$, see below), and the SNA is expected to be valid 
($\omega < m_\pi$) is delineated by thicker curves. The regions probed by
 the neutrino and axion emissivities are different, as indicated on
the plot.}
\label{ssig_1}
\end{figure}

\vskip -2 mm

Figure~\ref{ssig_1} shows the resulting ratio
$R_{\sigma}(\omega)$ for neutron matter at a range
of temperatures and a baryon density equal to the nuclear saturation
density. The results are plotted as a function of the dimensionless
ratio $\omega/T$ (note $S_\sigma(\omega)$ has significant strength
only for $\omega/T \lsim 15$).

The most striking feature of the results is that the one-pion-exchange
approximation significantly overestimates the rate for neutrino (or
axion) production. The large reduction in the response functions over
those obtained in the OPE approximation occurs for two
reasons. Firstly, one-pion exchange over-estimates the strength of the
$NN$ tensor force, and so even replacing the $V_{OPE}$ employed
previously with the full $V_{NN}$ would lead to a reduction in
$S_\sigma$. Secondly, the unitarity of our $NN$ T-matrix leads to an
$NN$ amplitude which is, in general, significantly smaller than that
found in the OPE approximation, and hence to a much-reduced
$S_\sigma$. That the OPE approximation does such a poor job in
describing the $NN$ dynamics should come as no surprise.  At the
energies of interest here the $NN$ interaction is intrinsically
non-perturbative, and so replacing the full $NN$ T-matrix by $V_{OPE}$
gives only a crude estimate of these rates. 

Fig.~\ref{ssig_1} may be used to infer where
the SNA breaks down. Recall that $S^{\rm ref}$ is constructed using
a hadronic tensor ${\cal H}_{ij}$ proportional to 
$1/\omega^2$. In fact this
part of the hadronic response is the only piece of the ${\cal H}_{ij}$
calculated in the SNA that can be trusted. Therefore, any
$\omega$-dependence of $R_\sigma(\omega)$ represents an effect in
$S_\sigma^{SNA}(\omega)$ coming from physics beyond the SNA. Viewing
Fig.~\ref{ssig_1} in this light, and considering the constraint
$\omega \lsim m_\pi$, implies that at $\rho=\rho_0$ the SNA
works well for $T \lsim 10$ MeV.

The suppression of the axial response seen in Fig.~\ref{ssig_1}
translates into a corresponding diminution of axion and neutrino
emissivities.  Let us define the ratio $R_{{\cal E}_\nu} \equiv {\cal
E}_{\nu\bar{\nu}}^{SNA}/{\cal E}_{\nu\bar{\nu}}^{OPE}$.  Here, ${\cal
E}_{\nu\bar{\nu}}^{OPE}$ is the emissivity found in the OPE
approximation, as calculated in Ref.~\cite{FM}. (A similar ratio of
axion emissivities is approximately equal to $R_{{\cal E}_\nu}$ in the
domain of validity of the SNA.) The ratio $R_{{\cal E}_\nu}$ is
displayed in the table below for a range of densitites and at
temperatures of 1 and 10 MeV.  In fact, the temperature and density
dependence of ${\cal E}_{\nu\bar{\nu}}^{SNA}$ is predominantly
determined by that of the nucleon equilibrium distribution functions
appearing in Eq.~(\ref{emiss}), but the ratio $R_{{\cal E}_\nu}$
changes significantly over the densities and temperatures considered
because ${\cal E}_{\nu \bar{\nu}}^{OPE}$ has much more variation with
$n_B$ and $T$.  We see that, for the range of conditions considered
here, the SNA gives a rate of emission of soft axial radiation which
is roughly a factor of four smaller than that given by the OPE
approximation.

\begin{center}
\begin{tabular}{|c|c|c|}
\hline
\, $n_B$ (fm$^{-3}$) \,&  
\, $R^{nn}_{{\cal E}_\nu}$ (1 MeV) \,& 
\, $R^{nn}_{{\cal E}_\nu}$ (10 MeV) \,\\
\hline \hline
0.08	&      0.29   &		   0.27		 \\
0.16	&      0.24   &            0.23		 \\
0.48	&      0.16   &		   0.16		 \\
\hline
\end{tabular} 
\end{center}
	
{\it Disclaimers:} As mentioned previously, our calculation makes no
attempt to include many-body effects. These will certainly be
important in some regimes of temperature and density. For instance, it
is claimed that in-medium modifications of the pion, attributed to the
many-body nature of the problem, strongly affect the emission
rates~\cite{VS}. Such effects are outside the scope of this work.
However, as stated earlier, we expect the LPM effect to strongly
reduce the response function~\cite{KV,RS,SD,Keil}. In particular, it
will suppress the emission of radiation with $\omega \lsim \gamma$,
where $\gamma$ is the nucleon quasi-particle width at the Fermi
surface.  This LPM-effect limit on the validity of our calculation is
indicated in Fig.~\ref{ssig_1}, with the value of $\gamma$ taken 
from Ref.~\cite{SD}. Figure \ref{ssig_1} suggests that LPM-suppression
will affect the emissivity if $T > 10$ MeV. Note 
that the LPM effect and the use of the SNA {\it both} 
significantly reduce the rate of emission of soft axial radiation.

Several microphysical ingredients play a role in the thermal evolution
of neutron stars, and so it is difficult to state precisely how the
results obtained in this work will affect observable aspects of
neutron star evolution. However, our results imply that $NN$
bremsstrahlung is less important during the star's infancy than
previously thought. In addition, the reduction of the axion emissivity
we have found will, presumably, weaken the axion-mass bound 
obtained from SN1987A. However, these comments are subject to
the caveat that through large regions of the infant neutron star
the temperature is high enough to invalidate the SNA.

On the other hand, for late-time cooling the temperature is small,
and our results are applicable everywhere in the
star. However, in this regime the modified URCA reaction, $nn
\rightarrow npe^-\bar{\nu_e}$, is significantly more efficient than
the pair process considered here~\cite{FM}.  In degenerate matter
this charged-current reaction does not produce soft radiation,
since the typical change in the energy of the $NN$
system is of order the electron chemical potential, i.e. about 100
MeV. Nevertheless, the results presented here suggest that large corrections to
the modified URCA rate calculated in the OPE approximation will occur
when a better model of the $NN$ amplitude is used.

{\it Conclusion:} Finally, we reiterate that none of these disclaimers
modify the two central conclusions of this paper: that the
soft-neutrino approximation gives a model-independent result for the
emissivity due to the reactions $NN \rightarrow
NN \nu \bar{\nu}$ and $NN \rightarrow NN a$; and that these
emissivities are much smaller than those found when one-pion exchange
is used as the $NN$ amplitude.

{\it Acknowledgements:} We are grateful to J.-W.~Chen and M.~J.~Savage for
useful discussions. 
We thank the U.~S. Department of Energy for its support under
contracts DOE/DE-FG03-97ER4014 and DOE/DE-FG06-90ER40561. C.~H. 
acknowledges the support of the Alexander von Humboldt foundation.

\end{document}